%% file: main.tex
\documentclass[
twocolumn,
amsfonts,
showpacs,
superscriptaddress,
amsmath,
amssymb,
aps,
longbibliography,
prl,
noeprint
]{revtex4-2}

\usepackage{graphicx}
\usepackage{bm}
\usepackage[dvipsnames]{xcolor}
\usepackage[english]{babel}
\usepackage{dsfont}
\usepackage{comment}
\usepackage[caption=false]{subfig}
\usepackage[colorlinks,%
            linkcolor=BrickRed,%
            citecolor=MidnightBlue,%
            urlcolor=MidnightBlue]{hyperref}
\usepackage{lipsum}

\newcounter{ls}

\newcommand{\bc}{\begin{center}}
\newcommand{\ec}{\end{center}}
\def\ba#1{\begin{array}{#1}\displaystyle}
\newcommand{\ea}{\end{array}}

\newcommand{\beq}{\begin{equation}}
\newcommand{\eeq}{\end{equation}}
\newcommand{\beqa}{\begin{eqnarray}}
\newcommand{\eeqa}{\end{eqnarray}}

\newcommand{\n}{\nonumber\\}
\newcommand{\bi}{\begin{itemize}}
\newcommand{\ei}{\end{itemize}}

\newcommand{\p}{\partial}

\newcommand{\ii}{{\rm i}}

\newcommand{\dd}{{\rm d}}

\newcommand{\eff}{{\rm eff}}

\DeclareMathOperator\erf{erf}

\begin{document}

\title{Hydrodynamic fluctuations of stochastic charged cellular automata}
\author{Takato Yoshimura}
\email{takato.yoshimura@physics.ox.ac.uk}
\affiliation{All Souls College, Oxford OX1 4AL, U.K.}
\affiliation{Rudolf Peierls Centre for Theoretical Physics, University of Oxford, 1 Keble Road, Oxford OX1 3NP, U.K.}

\author{\v{Z}iga Krajnik}
\email{ziga.krajnik@nyu.edu}
\affiliation{Department of Physics, New York University, 726 Broadway, New York, NY 10003, USA}
\begin{abstract}
We study charge fluctuations of a family of stochastic charged cellular automata away from the deterministic single-file limit and obtain the exact typical charge probability distributions, known to be anomalous, using hydrodynamics. The cellular automata considered are examples of linearly degenerate systems where two distinct mechanisms of diffusion, namely normal and convective diffusion, coexist. Our formalism, based on macroscopic fluctuation theory, allows us to describe current fluctuations stemming from these two diffusive processes, and we expect it to be applicable to generic linearly degenerate systems. The derived probability distributions match the exact microscopic result and numerical simulations.
\end{abstract}

\maketitle

{\it Introduction.}---%
Hydrodynamics is a powerful tool that allows us to describe the large-scale dynamics of systems in a universal way. Although hydrodynamics traditionally deals with the dynamics of mean values of observables, recently its scope has been extended to hydrodynamic fluctuations prescribed by correlation functions at hydrodynamic scales, culminating in the inception of macroscopic fluctuation theory (MFT) \cite{MFT} and its ballistic generalization, ballistic MFT (BMFT) \cite{BMFT,Doyon2023longrange}. Armed with experimental advances in probing the dynamics of cold atoms (e.g. quantum gas microscopy \cite{Scheie2021,Wei2022}), there has been also growing interest in experimental observations of hydrodynamic fluctuations, in part also fuelled by the discovery of anomalous spin current fluctuations in the isotropic Heisenberg chain \cite{Krajnik2022_1,Krajnik2024_1}, tantalizingly hinting at their surprising connection \cite{Das2019,Krajnik2020,Weiner2019,Dupont2020,Krajnik2020a,takeuchi2024} with the Kardar-Parisi-Zhang universality class \cite{Kardar1986,Corwin2012,Takeuchi2018}.

While hydrodynamic fluctuations in generic non-integrable and integrable systems have been rather well-understood so far, there is a class of systems whose hydrodynamic fluctuations are under-explored: linearly degenerate systems \cite{Lax1957,Tsarev_1991,Kamchatnov_2000,Bressan_book,Kuniba_2020}. Loosely speaking, these are non-integrable systems that are as close to integrability as possible in the sense that, in linearly degenerate systems, fluid modes do not self-interact. Such self-interactions are indeed also absent in integrable systems as trajectories of quasiparticles are only influenced by scatterings with other quasiparticles \cite{GHD1,Doyon_GHD_lecture}. The lack of self-interactions precludes the standard mechanism of superdiffusion \cite{Spohn2014, Hubner2025}, whereby leading corrections to ballistic transport become diffusive. Importantly, there are two mechanisms of diffusion, namely, normal and convective \cite{Medenjak2020} diffusion. The former is the standard source of diffusion in non-integrable systems whereas the latter, which often goes by the name of diffusion from convection, is an unconventional mechanism of diffusion and unique to linearly degenerate systems. While generically these two diffusion mechanisms coexist and affect typical fluctuations (hydrodynamic fluctuations at the diffusive scale) in linearly degenerate systems, it has remained unclear how to describe them consistently using hydrodynamic ideas. In this Letter, we address this issue by providing a framework based on MFT that enables us to systematically characterize typical fluctuations in generic linearly degenerate systems. It should also be emphasized that these systems are in fact not uncommon; recently some models with particle-hole symmetry such as Dirac fluids \cite{gopalakrishnan2024nongaussian} and quasi one-dimensional Hamiltonian systems \cite{McCulloch2025} have been shown to be linearly degenerate.

\begin{figure}
    \centering
    \includegraphics[width=\linewidth]{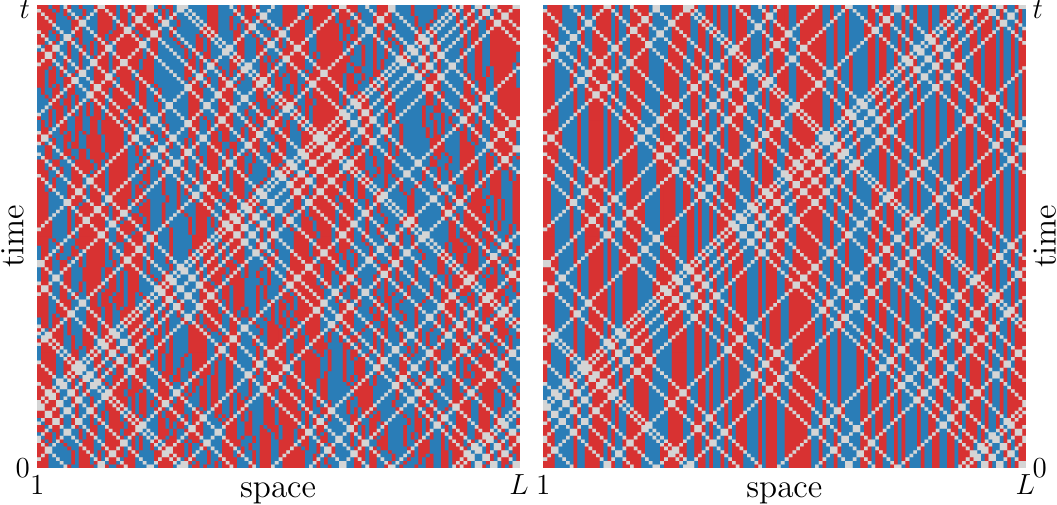}
    \caption{Many-body trajectories of the cellular automata starting from the same initial state for stochastic (left, $0<\Gamma<1$)  and (right, $\Gamma=0$) deterministic single-file dynamics. Vacancies (light grey) propagate freely while neighbouring positive/negative particles (red/blue) scatter stochastically with crossing probability $\Gamma$.}
    \label{fig:dynamics}
\end{figure}

As a concrete example to which we apply our framework, we focus on a family of stochastic charged cellular automata (SCCA). Their dynamics have been solved microscopically \cite{Medenjak_Klobas_Prosen_2017,Klobas_Medenjak_Prosen_2018,Medenjak2019}, including dynamical current fluctuations \cite{Krajnik2022,Krajnik_Schmidt_Pasquier_Prosen_Ilievski_2024,krajnik2025}. Strikingly, the SCCA exhibit anomalous fluctuations, manifested in a non-Gaussian typical charge current probability distribution. Recently in Ref.~\cite{Yoshimura2025} we have developed a MFT-based hydrodynamic approach to evaluate hydrodynamic fluctuations at the deterministic point of the SCCA where normal diffusion is absent, successfully reproducing earlier microscopic results \cite{Krajnik2022}. 
Our hydrodynamic framework, building on the approach developed in Ref.~\cite{Yoshimura2025}, recovers the typical charge fluctuations of the SCCA obtained recently in Ref.~\cite{krajnik2025} using methods of integrable combinatorics.

{\it Linearly degenerate systems.}---%
Consider a continuous one-dimensional quantum many-body system (generalization to higher dimensions, classical systems etc. is straightforward) with $N$ conserved charges $\hat{Q}_i=\int\dd x\,\hat{q}_i(x)$, where the density $\hat{q}_i(x)$ satisfies the continuity equation $\p_t\hat{q}_i(x,t)+\p_x\hat{\jmath}_i(x,t)=0$ with the associated current $\hat{\jmath}_i(x)$. Since linear degeneracy is a concept associated to hyperbolic systems of conservation laws, we assume ballistic transport in the system. Namely, we assume that the flux jacobian $\mathsf{A}_i^{~j}[\underline{\langle\hat{q}\rangle}]=\p \langle\hat{\jmath}_i\rangle/\p\langle\hat{q}_j\rangle$ (here the underline indicates that the quantity is a vector), where the average $\langle\cdots\rangle$ is with respect to a generic generalized Gibbs ensemble (GGE), 
is a nonzero matrix. The corresponding Euler hydrodynamic equation for the Euler scale average densities $\mathsf{q}_i(x,t)=\lim_{\tau\to\infty}\langle \hat{q}_i(\tau x,\tau t)\rangle$ and currents $\mathsf{j}_i(x,t)=\lim_{\tau\to\infty}\langle \hat{\jmath}_i(\tau x,\tau t)\rangle$ reads $\p_t\mathsf{q}_i(x,t)+\p_x\mathsf{j}_i(x,t)=0$, which can also be expressed as
\begin{equation}\label{eq:euler_hydro}
    \p_t\mathsf{q}_i(x,t)+\mathsf{A}_i^{~j}[\underline{\mathsf{q}}(x,t)]\p_x\mathsf{q}_j(x,t)=0.
\end{equation}
While it is natural to write down hydrodynamic equations for physical charges, a more convenient basis of hydrodynamic modes is that of {\it normal modes}. Diagonalizing the the flux jacobian as $\mathsf{R}\mathsf{A}\mathsf{R}^{-1}=\mathrm{diag}\,\underline{v^\eff}$ where $v^\eff_i$ are the eigenvalues of $\mathsf{A}$, the normal modes $n_i$ are defined via the transition matrix $\mathsf{R}_i^{~j}$ by the relation $\mathsf{R}_i^{~j}=\p n_i/\p\mathsf{q}_j$, in terms of which Eq.~\eqref{eq:euler_hydro} becomes
\begin{equation}\label{eq:normal_hydro}
    \p_tn_i(x,t)+v^\eff_i(x,t)\p_xn_i(x,t)=0.
\end{equation}
Normal modes also elucidate the physical meaning of linear degeneracy -- a hydrodynamic system Eq.~\eqref{eq:normal_hydro} is linearly degenerate precisely when {\it every normal mode $n_i$ is non self-interacting, $\p v_i^\eff/\p n_i=0$}, that is, the dynamics of a normal mode is not influenced by the density of the mode itself. This condition is non-generic and self-interaction is known to be the source of superdiffusion in one dimensional continuum systems \cite{Spohn2014}. A notable class of linearly degenerate systems are integrable systems where normal modes are given by quasi-particles, whose trajectories are changed only by collisions with other quasi-particles, hence the absence of self-interaction.

Since superdiffusion is avoided due to the absence of self-interactions in linearly degenerate systems \cite{self_coupling}, leading corrections to ballistic transport are expected to be diffusive. In particular, there are generically two sources of diffusion: normal and convective diffusion. Normal diffusion is an ubiquitous source of diffusion where a slow mode constantly undergoes collisions with fast modes, giving rise to a jiggling movement of the slow mode. Such movement is characterized by Fick's law for the slow mode's current while the course-grained fast modes are modelled as an additive white noise to ensure the fluctuation-dissipation relation.

Convective diffusion, also known as diffusion from convection \cite{Medenjak2020}, has a completely different origin. In linearly degenerate systems hydrodynamic modes interact in a way that propagates initial fluctuations ballistically, and as a result, initial fluctuations survive even at finite times, contributing to the diffusive scale. This can be best appreciated in integrable systems: in these systems quasi-particles scatter purely elastically, indicating that hydrodynamic fluctuations are transferred without being dissipated.

In what follows we focus on a specific class of linearly degenerate systems called stochastic charged cellular automata (SCCA). 
While our choice facilitates analytic calculations, we expect the developed formalism and results are applicable to any linearly degenerate system.

{\it The model.}---%
We consider a family of SCCA \cite{Klobas_Medenjak_Prosen_2018} on a one-dimensional lattice of $L \in 2\mathbb{N}$ sites. The lattice configuration at fixed time $t \in \mathbb{Z}$ is given by a string ${\bf s}^t \equiv s^t_{1}s^t_{2} \ldots s^t_L$
of symbols $s^t_x \in \{\emptyset, +, -\}$, which correspond to vacancies and positively or negatively charged particles respectively.
Symbol dynamics are locally given by a one-parameter stochastic two-body map $(s_L', s_R') = \Phi(s_L, s_R)$ which encodes the following dynamical rules $(\emptyset, \emptyset) \rightarrow (\emptyset, \emptyset)$, $(\emptyset, c) \leftrightarrow (c, \emptyset)$ while $(c, c') \to (c', c)$ with crossing probability  $ 0 \leq \Gamma \leq 1$ and $(c, c') \to (c, c')$ with reflection probability $\overline \Gamma \equiv 1-\Gamma$ where $c,c' \in \{-,+\}$.
The many-body dynamics is realized as a `brickwork' circuit obtained by imposing periodic boundary conditions, $s^t_{x} = s^t_{L+x}$, and coupling alternating sites as ${\bf s}^{2t+2} = \Phi^{\rm odd}({\bf s}^{2t+1}) $ and  $ {\bf s}^{2t+1} = \Phi^{\rm even}({\bf s}^{2t})$, where \begin{equation}
\Phi^{\rm odd}=\prod_{x=1}^{L/2}\Phi_{2x-1,2x}, \qquad  \Phi^{\rm even}=\prod_{x=1}^{L/2}\Phi_{2x,2x+1}
\end{equation}
and $\Phi_{x,x+1}$ acts non-trivially only on sites $x$ and $x+1$, see Fig.~\ref{fig:dynamics} for representative many-body trajectories. For $\Gamma=0$ the dynamics is deterministic and corresponds to \emph{single-file} dynamics \cite{Medenjak_Klobas_Prosen_2017,Medenjak2019,Krajnik2022,Krajnik_Schmidt_Pasquier_Prosen_Ilievski_2024,krajnik2024singlefile} where the only source of diffusive scale dynamics is convective diffusion \cite{Yoshimura2025} while for $\Gamma = 1$ the dynamics is free with no diffusive dynamics.

We restrict our study to a minimal subset of conserved charges, comprising the number of right and left moving particles $\hat{Q}_\pm$ and the total charge $\hat{Q}_c$, all of which are given as sums of local densities $\hat{Q}_i({\bf s}) = \sum_{x=1}^{L} \hat{q}^{i}_{x, t}({\bf s})$ where $i \in  \mathcal{I} \equiv \{+, -, c\}$ and $\hat \bullet$ indicates a microscopic quantity, see SM \cite{SM} for details on charge densities and discrete space-time continuity relations.
The SCCA are super-integrable and support an exponential (in $L$) number of local conserved quantities \cite{Gombor2022} which are invariant under charge conjugation due to the underlying free particle dynamics.
Note however that, for $\Gamma \neq 1$, only the total charge is odd under charge conjugation, while for $\Gamma = 1$ there exist exponentially many conserved quantities odd under charge conjugation.

We consider generalized Gibbs ensembles (GGEs) of the form $\mathbb{P}({\bf s}) = e^{\beta^i \hat{Q}_i({\bf s})}/\sum_{\bf s'} e^{\beta^i \hat{Q}_i({\bf s'})},$ where repeated upper and lower indices indicate summation.
The average of an observable $o({\bf s})$ is accordingly $\langle o \rangle = \sum_{{\bf s}} \mathbb{P}({\bf s})\, o({\bf s}).$ Identifying the ensemble parameters as 
$\rho_\pm = [1 + e^{-\beta_\pm}/(2 \cosh \beta_c)]^{-1}, b =  \tanh \beta_c$,
the ensemble factorizes
$\mathbb{P}({\bf s}) = \prod_{x=1}^{L/2}p_+(s_{2x}) p_-(s_{2x+1})$
in terms of normalized one-site measures $p_-(\pm) = \rho_- \frac{1\pm b}{2}$, $p_+(\pm) = \rho_+ \frac{1\pm b}{2}$, $p_\pm(\emptyset) = \overline \rho_\pm$ with $0 \leq \rho_\pm \leq 1$ and $\overline \rho_\pm \equiv 1 - \rho_\pm$ the densities of right/left movers and vacancies on right/left running diagonals, respectively.

The dynamics of a GGE comprising the three charges $\hat{Q}_\pm$ and $\hat{Q}_c$ is closed, i.e.~at the Euler scale their densities evolve without reference to the dynamics of other local conserved charged. This is reflected in the structure of the three-mode Euler hydrodynamics obtained by evaluating the flux jacobian \cite{Yoshimura2025}
\begin{equation}\label{eq:flux_jacobian}
    [\mathsf{A}[\underline{\rho}]]_i^{~j}=\frac{\p j_i[\underline{\rho}]}{\p\rho_j}=\begin{pmatrix}
    1 & 0&0\\
    0&-1&0\\
    b\rho_-/\rho&-b\rho_+/\rho& v
    \end{pmatrix}
\end{equation}
 where the three modes are labelled as $\rho_1=\rho_+/2$, $\rho_2=\rho_-/2$, $\rho_3=\rho_c$, and  $v=p/\rho$ is the charge velocity with $p= \rho_1 -\rho_2$ and $\rho = \rho_1 + \rho_2$. Here $j_i[\underline{\rho}]$ is the average current defined via the continuity equation $\p_t\varrho_i+\p_xj_i[\underline{\varrho}]=0$ where $\varrho_i$ is a space-time dependent density. While the dynamics of the left and right movers are trivial $\p_t \varrho_\pm\pm\p_x \varrho_\pm =0$, charge transport is nontrivially affected by particle dynamics; the Euler equation accordingly becomes
\begin{equation}\label{eq:charge_hydro}
    \p_t\varrho_c+\p_x(v\varrho_c)=0.
\end{equation}
Since the Euler hydrodynamic equations of the SCCA are manifestly linearly degenerate, corrections to the ballistic transport in the SCCA are generically diffusive as noted previously. Both normal and convective diffusion are present in the stochastic dynamics (for $0<\Gamma<1$), but the former disappears in the single-file limit $\Gamma=0$, which is the case treated in our previous work Ref.~\cite{Yoshimura2025}. In what follows we characterize the typical charge current probability distribution when both diffusion mechanisms coexist using an idea based on MFT.

{\it Typical current fluctuations from MFT.}---%
A standard object used to probe the statistics of the time-integrated charge current $\hat{J}(t)=\int_0^t\dd t'\,\hat{\jmath}_{c}(0,t')$ is the generating function $\langle e^{\ii \lambda \hat{J}(t)}\rangle=\int\dd\lambda\,e^{\ii\lambda J}\mathcal{P}(J|t)$ of the probability distribution $\mathcal{P}(J|t)$. We are presently interested in the typical probability distribution for currents of magnitude $\hat{J}(t)\sim t^{1/{2z}}$, where the dynamical exponent $z$ is set by the variance $\langle \hat{J}^2\rangle^c(t)\sim t^{1/{z}}$. We accordingly define the normalized typical probability distribution $\mathcal{P}_\mathrm{typ}(j)=\lim_{t\to\infty}t^{1/2z}\mathcal{P}(jt^{1/2z}|t)$. On the other hand, large current fluctuations of order $t$, i.e. $\hat{J}(t)\sim t$, satisfy a large deviation principle \cite{Touchette_2009} of the form $\mathcal{P}(jt|t)\asymp e^{-tI(j)}$, where $I(j)$ is the rate function.

The exact microscopic probability distribution of the charge current in the single-file limit  $\Gamma=0$ of the SCCA was first obtained analytically using combinatorial techniques in Ref.~\cite{Krajnik2022} and later reproduced in Ref.~\cite{Yoshimura2025} based on a hydrodynamic approach we also employ here. While these studies focused on the deterministic case, more recently the full probability distribution for generic $\Gamma>0$ was microscopically obtained by mapping the charge dynamics to the stochastic six vertex model in Ref.~\cite{krajnik2025}. 

The fundamental tenet of MFT (and its ballistic generalization BMFT) is that for large fluctuations $\hat{J}(t)\sim t$, the generating function $\langle e^{\ii \lambda \hat{J}(t)}\rangle$ is fully determined by the saddle point of a path integral over fluctuating hydrodynamic fields. This means that MFT allows us to evaluate the generating function and in general any physical quantities controlled by \emph{large} fluctuations using solely hydrodynamic data. In our recent work Ref.~\cite{Yoshimura2025}, we have applied a similar idea to typical fluctuations and successfully reproduced the exact \emph{typical} charge probability distribution for the SCCA at the deterministic point. As we demonstrate, an extension of the approach applies to any linearly degenerate system, in which case diffusion has two distinct origins. We first outline the formalism for generic linearly degenerate systems and then apply it to the SCCA.

Let $q_i(x,t)=\hat{q}_i(\tau^{1/z}x,\tau t)$ be the fluctuating hydrodynamic fields at the typical scale, where $\tau \gg 1$ is the time over which we evaluate the current statistics. Note that the dynamical exponent can vary depending on the chosen charge (e.g. $z=2$ for charge transport while $z=1$ for left/right movers in the SCCA). At the typical scale the fields are initially Gaussian-distributed by the central limit theorem and can be decomposed as
\begin{equation}
    q_i(x,0)\simeq \mathsf{q}_i+\tau^{-1/2z}\delta q_i(x,0),
\end{equation}
where $\mathsf{q}_i$ is the initial average density and $\delta q_i(x,0)$ is a white noise whose variance is set by the susceptibility matrix $\mathsf{C}_{ij}=\p \mathsf{q}_i/\p\beta^j$. For simplicity we assume that the initial GGE is chosen such that $\mathsf{j}_i=0$ for the charge whose statistics we study. With this initial condition, we claim that the fluctuating fields $q_i(x,t)$ evolve according to the following fluctuating hydrodynamic equation
\begin{equation}\label{eq:mft_hydro}
     \p_tq_i+\tau^{1/z}\mathsf{A}_i^{~j}[\underline{q}]\p_xq_j+\p_x\left(-\frac{1}{2}\mathsf{D}_c^{~i}\p_xq_i+\eta_i\right)=0,
\end{equation}
where $\mathsf{D}_i^{~j}$ is the {\it projected} diffusion matrix and $\eta_i$ is a white noise $\overline{\eta_i(x,t)\eta_j(x',t')}=\tau^{-1/2}\sigma_{ij}\delta(x-x')\delta(t-t')$ with the noise strength set by the projected Onsager matrix $\sigma_{ij}=\int\dd t\dd x\overline{\langle \hat{\jmath}_i(x,t)\hat{\jmath}_j(0,0)\rangle^c}-\sigma_{ij}^\mathrm{conv}$. Here $\sigma_{ij}^\mathrm{conv}$ refers to the convective contribution to the Onsager matrix obtained in Ref.~\cite{Medenjak2020} (see End Matter for its explicit expression), which stems convective diffusion, and the projected diffusion matrix is accordingly related to the projected Onsager matrix via the fluctuation-dissipation relation $\mathsf{D}\mathsf{C}=\sigma$, both of which are evaluated with respect to the initial GGE. 

\begin{figure}
\includegraphics[width=\linewidth]{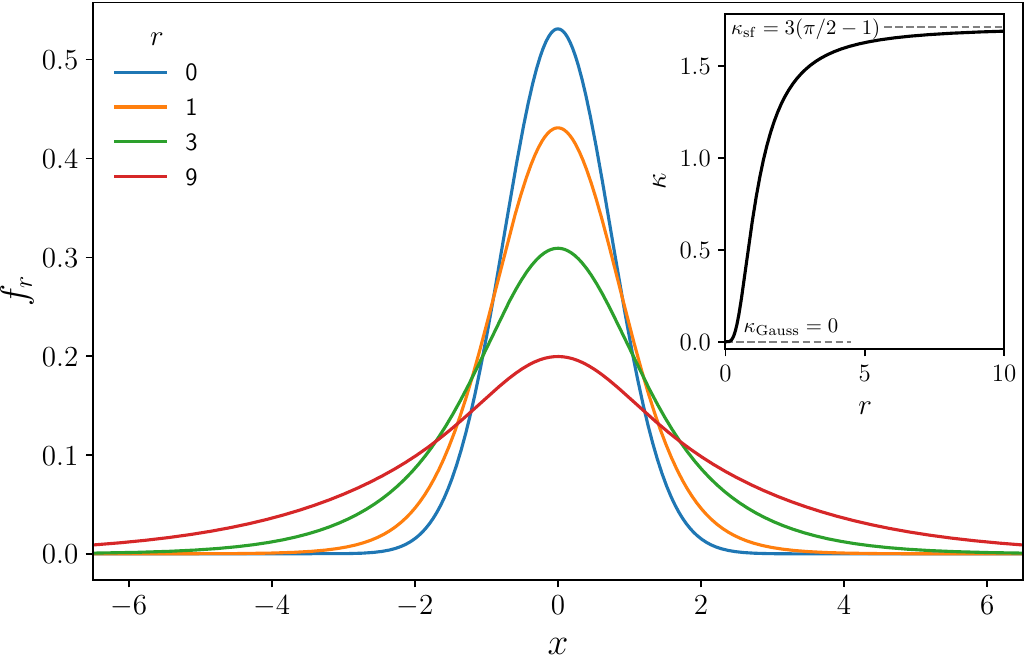}
\caption{(main panel) Scaling probability distribution \eqref{eq:typ_prob} (coloured curves) for different ratios $r$. The characteristic non-analyticity of the single-file distribution for $r\to \infty$ is due to the divergent rescaling as $\omega \to 0$.
(Inset) Excess kurtosis $\kappa$ \eqref{kurtosis} of the scaling distribution (black curve) interpolates between the Gaussian value $\kappa_{\rm Gauss} = 0$ and the single-file value $\kappa_{\rm sf} = 3(\pi/2-1)$ (dashed gray lines) with increasing $r$.
}
\label{fig:scaling_pdf}
\end{figure}

Some comments are in order. First, the hydrodynamic equation Eq.~\eqref{eq:mft_hydro} reduces to the Euler equation in the absence of normal diffusion, which described typical charge fluctuations in the single-file limit of the SCCA \cite{Yoshimura2025}, consistent with the appearance of the projected diffusion matrix. Second, Eq.~\eqref{eq:mft_hydro} should not be mistaken for the standard hydrodynamic equation for the averages that takes a different form: in Ref.~\cite{Doyon_noise_2025}, the full hydrodynamic equation for linearly degenerate systems up to diffusive corrections is obtained by generalizing the hydrodynamic equation reported in Ref.~\cite{Hubner2025}, which did not include normal diffusion.


{\it Typical charge fluctuations of the SCCA.}---%
We now apply Eq.~\eqref{eq:mft_hydro} to charge fluctuations in the SCCA for ensembles with vanishing total charge where the charge dynamical exponent is $z=2$.  We first determine the projected diffusion and Onsager matrices which are informed by microscopics. Since normal diffusion is unaffected by the underlying movement of particles, the relevant elements of the diffusion matrix vanish $\mathsf{D}_c^{~\pm}=0$, leaving only $D=\mathsf{D}_c^{~c}$ to be determined. Using the techniques developed in Ref.~\cite{krajnik2025}, in SM \cite{SM} we explicitly evaluate the asymptotic behaviour of the charge dynamical structure factor which, on the diffusive scale, turns out to be Gaussian with variance $(1-\overline{\Gamma}\varrho)/\overline{\Gamma} \varrho$. The charge diffusion coefficient stemming from diffusion from convection is $(1-\varrho)/\varrho$ and the projected charge diffusion coefficient thus becomes $D=(1-\overline{\Gamma}\varrho)/\overline{\Gamma}\varrho-(1-\varrho)/\varrho=\gamma/\rho$, where $\gamma=\Gamma/\overline \Gamma$, which also gives $\sigma_{cc}=\gamma$.  The hydrodynamic equation for the fluctuating charge density thus reads
\begin{equation}\label{eq:mft_hydro_scca}
    \p_t\varrho_c+\tau^{1/2}\p_x(v\varrho_c)+\p_x\left(-\frac{1}{2}D\p_x\varrho_i+\eta_c\right)=0
\end{equation}
with $\overline{\eta_c(x,t)\eta_c(x',t')}=\tau^{-1/2}D\rho\delta(x-x')\delta(t-t')$. 
To solve Eq.~\eqref{eq:mft_hydro_scca} at large $\tau$, we first note that stochasticity does not influence the dynamics of left/right movers and their dynamics are still given by $\p_t\varrho_\pm\pm\tau^{1/2}\p_x\varrho_\pm=0$, which is solved by $\varrho_\pm(x,t)=\rho+\tau^{-1/4}\delta\varrho_\pm(x\mp\sqrt{\tau}t,0)$. The charge density then satisfies $\varrho_c(x,t)\simeq \rho b(x,t)$ at large $\tau$, allowing us to obtain the hydrodynamic equation for $b(x,t)$ from Eq.~\eqref{eq:mft_hydro_scca}. The resulting equation is solved by a convolution (see End Matter for the derivation), which in turn gives the full solution of $\varrho_c(x,t)$ at large $\tau$
\begin{align}\label{eq:hydro_solution}
    \varrho_c(x,t)&\simeq\int\dd y\,G_t(x-X_t-y)\delta\varrho_c(y,0) \\
    &+\int_0^t\dd s\int\dd y\, G_{t-s}(x-X_t-y)\partial_y\eta(y,s), \nonumber
\end{align}
where $G_t(x)=e^{-x^2/(2Dt)}/\sqrt{2\pi Dt}$ is the Gaussian kernel and $X_t$ is the characteristics of $b(x,t)$ that satisfies $\dd X_t/\dd t=v(X_t,t))$ with $X_0=0$. Using the solution \eqref{eq:hydro_solution}, the time-integrated current $\hat{J}(\tau)=\sqrt{\tau}\int_0^\infty\dd x\,(\delta\varrho_c(x,1)-\delta\varrho_c(x,0))$ is also expressed in terms of the initial density fluctuations $\delta\varrho_c(x,0)$ and the noise. 
We can now write down an MFT-like path-integral for the generating function. In End Matter, we carry out the path-integral and show that it can be cast as a single-variable integral
\begin{align} \label{eq:MGF}
    \overline{\langle e^{\lambda\hat{J}_c(\tau)}\rangle}&\simeq \frac{1}{Z}\int\mathcal{D}\underline{\delta\varrho}\,\overline{e^{-\mathsf{C}^{ij}\int\dd x\,\delta\varrho_i(x,0)\delta\varrho_j(x,0)/2}e^{\lambda\hat{J}_c(\tau)}} \n
    &=\int_{\mathbb{R}} \frac{\dd x}{\sqrt{2\pi\Delta^2}}e^{-x^2/2\Delta^2 + (\tau/2)^{1/2} \lambda^2 \omega g^2(x/\omega)/2}
\end{align}
where $g^2(x) = x\, {\rm erf}\, x + \pi^{-1/2} e^{-x^2}$ while $\Delta^2=2\rho \overline \rho$ (with $\overline \rho = 1-\rho)$ and $\omega^2= 4 \gamma \rho$. The corresponding typical probability distribution assumes a one-parameter scaling form $\mathcal{P}_{\rm typ}(j) = \omega^{-1/2} f_r(j/\omega^{1/2})$ controlled by the ratio  between convective and normal diffusion 
$r \equiv \Delta/\omega = \overline \rho/2\gamma$ 
\begin{equation} \label{eq:typ_prob}
    f_r(x) =  \int_{\mathbb{R}} \frac{\dd y}{2\pi g(ry)}\, e^{-y^2/2 -x^2/2g^2(ry)}.
\end{equation}
The scaling distribution \eqref{eq:typ_prob} (see Fig.~\ref{fig:scaling_pdf}) coincides with the microscopic result obtained in Ref.~\cite{krajnik2025} using integrability (apart from a numerical factor due to different conventions used) and reduces to the single-file result \cite{Krajnik2022,Yoshimura2025} for $D\to0$.

The contributions of normal and convective diffusion to the dynamical structure factor are additive and therefore not readily distinguishable. On the other hand, knowledge of the distribution \eqref{eq:typ_prob} allows one to quantify the relative strengths of the two sources of diffusion by determining the ratio $r$, e.g.~by measuring the excess kurtosis $\kappa = \lim_{t \to \infty} \langle \hat J^4 \rangle^c(t) / [\langle \hat J^2 \rangle^c(t)]^2 - 3$, see inset of Fig.~\ref{fig:scaling_pdf} and Eq.~\eqref{kurtosis} of End Matter for its explicit form. We note that kurtosis measurements in quantum integrable systems are becoming feasible \cite{Rosenberg_2024}.

{\it Conclusions.}---%
We have developed a hydrodynamic framework based on MFT to describe fluctuations in linearly degenerate systems. Notably, two disparate mechanisms, namely normal and convective diffusion, coexist in generic linearly degenerate systems, and our approach allows us to treat these diffusion processes consistently. To check the validity of our method, we studied a family of stochastic charged cellular automata (SCCA) and evaluated the typical charge current probability distribution, which is known to be anomalous (non-Gaussian). The resulting distribution Eq.~\eqref{eq:typ_prob} coincides with the microscopic result obtained recently which in turn agrees with numerical simulations \cite{krajnik2025}.

There are several directions that deserve further investigations. While linear degeneracy has a simple definition in terms of fluid modes, it remains largely unclear how the hydrodynamic condition translates to microscopic models. In a similar vein, is it be possible to systematically break integrability \cite{Prelovsek2,Pawlowski_2025} in a way that linearly degeneracy is kept intact? Answering these questions would be of great interest and would give invaluable insights on the classification of non-integrable and integrable systems. It would also be worthwhile to formulate linearly degenerate hydrodynamics up to diffusive corrections in a more mathematically rigorous way.

{\it Acknowledgements.}---%
We thank Benjamin Doyon for useful discussions and Vincent Pasquier for comments on the manuscript.
\v{Z}K is supported by the Simons Foundation as a Junior Fellow of the Simons Society of Fellows (1141511).

{\it Note added.}---%
We recently became aware of a forthcoming related and independent work by Benjamin Doyon \cite{Doyon_noise_2025}. In Ref.~\cite{Doyon_noise_2025} the same projected diffusion matrix is obtained for linearly degenerate systems by evaluating the corrections to Euler-scale correlation functions, while in this Letter we directly take the diffusive scaling and evaluate current fluctuations and correlation functions.
\appendix
\section{End Matter}
\subsection{Solution of the hydrodynamic equation for $\varrho_c(x,t)$}
According to Eq.~\eqref{eq:mft_hydro_scca}, the equation for the bias field $b(x,t)$ reads
\begin{equation}\label{hydro_normaldiff_2}
    \p_tb+\sqrt{\tau}v\p_xb=\frac{1}{2}D\p_x^2b+\rho^{-1}\p_x\eta_c,
\end{equation}
which is readily solved by a convolution. Let us first notice that the solution of the equation $\p_tb+\sqrt{\tau}v\p_xb=0$, which we denote by $\tilde{b}(x,t)$, can be obtained in terms of characteristics. Defining $\dd X_{x,t}/\dd t=\sqrt{\tau}v(X_{x,t},t)$ where $X_{x,0}=x$, it follows that $\tilde{b}(X_{x,t},t)$ is constant in time, i.e. $\tilde{b}(X_{x,t},t)=\tilde{b}(x,0)$. In fact, when $\tau$ is large we can neglect the $x$-dependence of $X_{x,t}$ because
\begin{align}
    \frac{\dd X_{x,t}}{\dd t}&\simeq\frac{\tau^{1/4}(\delta\varrho_+(X_{x,t}-\sqrt{\tau}t,0)-\delta\varrho_-(X_{x,t}+\sqrt{\tau}t,0))}{2\rho}\n
    &\simeq \tau^{1/4}\frac{\delta\varrho_+(-\sqrt{\tau}t,0)-\delta\varrho_-(\sqrt{\tau}t,0)}{2\rho}.
\end{align}
We can thus approximate $X_{x,t}$ as $X_{x,t}\simeq x+X_t$ and have $\tilde{b}(x,t)=\tilde{b}(x-X_t,0)\simeq \tau^{-1/4}\delta\tilde{\varrho}_c(x-X_t,0)/\rho$. Then the solution of Eq.~\eqref{hydro_normaldiff_2} at large $\tau$ is given by
\begin{align}
     b(x,t)&\simeq \int\dd y\,G_t(x-y)\tilde{b}(y,t)\n
     &\quad+\rho^{-1}\int_0^t\dd s\int\dd y G_{t-s}(x-X_t-y)\partial_y\eta(y,s) \n
     &=\int\dd y\,G_t(x-X_t-y)\tilde{b}(y,0)\n
     &\quad+\rho^{-1}\int_0^t\dd s\int\dd y G_{t-s}(x-X_t-y)\partial_y\eta(y,s),
\end{align}
which yields Eq.~\eqref{eq:hydro_solution}.
\subsection{Projected diffusion matrix}
Here we shall show that the diffusion matrix used for the Fick's law in Eq.~\eqref{eq:mft_hydro_scca} has to be the projected one. In terms of the fluctuating charge field at the typical scale $\varrho_c(x,t)$, the total Onsager matrix is defined by
\begin{equation}
    \sigma^\mathrm{tot}_{ij}=\lim_{\tau\to\infty}\sqrt{\tau}\int\dd x\int_0^1\dd t\int_0^1\dd t'\langle j_i(x,t)j_j(0,t')\rangle^c.
\end{equation}
To compute this, we first note that
\begin{align}
    \int_0^1\dd t\,j_c(x,t)
    &=\int_x^\infty\dd y\,(\varrho_c(y,\tau)-\varrho_c(y,0)) \\
    &=\tau^{-1/4}\sqrt{\rho}\int\dd y\,\mu(x,y)\delta n_c(y,0)\n
    &\quad+\int_0^1\dd s\int\dd y\,G_{1-s}(x-y-X_1)\eta(y,s), \nonumber
\end{align}
where $\mu(x,y)=\mathrm{erfc}((x-y-X_1)/\sqrt{2D})/2-\Theta(y-x)$. Using this expression, we then have
\begin{align}
    &\int\dd x\int_0^1\dd t\int_0^1\dd t'\overline{\langle j_c(x,t)j_c(0,t')\rangle^c}\n
    &=\tau^{-1/2}\rho\left(\int\dd x\dd y \dd z\,\langle \mu(x,y)\mu(0,z)\rangle_{\delta n_1,\delta n_2}+D\right),
\end{align}
where $\langle\cdots\rangle_{\delta n_1,\delta n_2}$ indicates the (path-integral) averaging over $\delta n_1$ and $\delta n_2$ where we used that both $\delta n_c(x,0)$ and $\eta(x,t)$ are Gaussian distributed with mean zero and $\int \dd x\,G_t(x)=1$. A lengthy but straightforward calculation now shows that
\begin{align}
    &\tau^{-1/2}\rho\int\dd y\int \dd z\, \mu(x,y)\mu(0,z)\\
    &=\tau^{-1/2}\rho|X_1|^2=\tau^{-1}(1-\rho)\left(\int_0^{\sqrt{\tau}}\dd x\,\delta\bar{n}(x,0)\right)^2, \nonumber
\end{align}
where $\delta \bar{n}(x,0)=(\delta n_1(x,0)-\delta n_2(-x,0))/\sqrt{2}$. Performing the path-integral over $\delta n_1$ and $\delta n_2$, we obtain
\begin{equation}
    \tau^{-1/2}\rho\int\dd x\dd y\dd z\, \langle\mu(x,y)\mu(0,z)\rangle_{\delta n_1,\delta n_2}=\tau^{-1/2}\overline \rho.
\end{equation}
Combining these, we thus end up with
\begin{equation}
    \sigma^\mathrm{tot}_{cc}=\overline \rho + D\rho.
\end{equation}
Since $\sigma^\mathrm{conv}_{cc}=\overline \rho$ and $\sigma_{cc}=D\rho$, we establish that $\sigma_{cc}$ is indeed the projected Onsager matrix $\sigma_{cc}=\sigma^\mathrm{tot}_{cc}-\sigma^\mathrm{conv}_{cc}$. It should be stressed that the above derivation can be readily generalizable to any linearly degenerate systems. We thus expect that the projected diffusion matrix controls the strength of normal diffusion via the Fick's law in generic linearly degenerate systems.

\subsection{Typical charge probability distribution}
Here we compute the generating function $\overline{\langle e^{\lambda\hat{J}(\tau)}\rangle}$ using a path-integral. We first carry out the path-integral for $\delta n_c$ and average over the noise $\eta_c$, which gives
\begin{align}
    \overline{\langle e^{\lambda\hat{J}_c(\tau)}\rangle}&\simeq \left<e^{\frac{1}{2}\lambda^2\rho\tau^{1/2}\left(|X_1|+\sqrt{\frac{2D}{\pi}}p\left(\frac{|X_1|}{\sqrt{2D}}\right)\right)}\right>_{\delta n_1,\delta n_2} \n
    & \left<e^{\frac{1}{2}\lambda^2\rho\tau^{1/2}\left(|X_1|+\sqrt{\frac{2D}{\pi}}p\left(\frac{|X_1|}{\sqrt{2D}}\right)\right)}\right>_{\delta\bar{n}}
\end{align}
Since $X_1=\tau^{-1/4}\sqrt{\overline \rho/\rho}\int_0^{\sqrt{\tau}}\dd x\,\delta \bar{n}(x,0)$ is still Gaussian-distributed, this path integral can be reduced to a one-dimensional integral. Namely, we have that
\begin{equation}\label{eq:identity}
    \int\mathcal{D}\phi\,e^{-\int\dd x\frac{
    \phi(x)^2}{2}+\mathsf{f}(\int\dd x\,\mathsf{g}(x)\phi(x))}=\int \dd X\,e^{-X^2/2+\mathsf{f}(\hat{\mathsf{g}}X)},
\end{equation}
for generic functions $\mathsf{f}(x)$ and $\mathsf{g}(x)$ where $\hat{\mathsf{g}}=\sqrt{\int\dd x\,\mathsf{g}(x)^2}$, which follows from the following discrete version of the identity \eqref{eq:identity}
\begin{equation}
    \int\frac{\dd^nx}{(2\pi)^{n/2}}e^{-\frac{1}{2}\underline{x}\cdot\underline{x}}e^{\mathsf{f}(\underline{\mathsf{g}}\cdot\underline{x})}=\int\frac{\dd x}{\sqrt{2\pi}}e^{-\frac{1}{2}x^2}e^{f(\hat{\mathsf{g}}x)},
\end{equation}
where $\hat{\mathsf{g}}=\sqrt{\underline{\mathsf{g}}\cdot\underline{\mathsf{g}}}$ and $\underline{\mathsf{g}}\in\mathbb{R}^n$. The continuum limit of this then yields Eq.~\eqref{eq:identity}, which in turn allows us to obtain Eq.~\eqref{eq:MGF}. Inverting the Laplace transform, the typical probability distribution  $\mathcal{P}_\mathrm{typ}(j)$ is then given by  Eq.~\eqref{eq:typ_prob} whose second moment reads
\begin{equation}
m_2 = \int_{\mathbb{R}} \dd x\,  x^2 P_r(x) = \sqrt{(1 + 2r^2)/\pi}.
\end{equation}
We also compute the fourth moment
\begin{equation}
m_4 = \frac{3}{\pi} \frac{1+6r^2}{1+2r^2} \frac{1}{\sqrt{1+4r^2}} + 3r^2\int_{\mathbb{R}} \tfrac{\dd y}{\sqrt{2\pi}} \, y^2 \erf^2(r y) e^{-y^2/2},
\end{equation}
whence we extract the excess kurtosis  (see inset of Fig.~\ref{fig:scaling_pdf})
\begin{align}
3 + \kappa(r) = m_4/m_2^2 &= \frac{3}{1+2r^2} \left[ \frac{1+6r^2}{1+2r^2}\frac{1}{\sqrt{1+4r^2}} \right. \label{kurtosis} \\
&\left. +\ \pi r^2 \int_{\mathbb{R}} \tfrac{\dd y}{\sqrt{2\pi}}\, y^2 \erf^2(r y)e^{-y^2/2}\right], \nonumber
\end{align}
with the limits $\kappa(0) = 0$ and $\kappa(\infty) = 3(\pi/2-1)$ which agree with Gaussian and single-file values respectively. Note that the kurtosis is unchanged by rescaling so that \eqref{kurtosis} holds for the typical distribution even for $r \to \infty$ despite the involved divergent rescaling as $\omega \to 0$.

\bibliography{bib.bib}
\clearpage
\onecolumngrid
\input{sm1.tex}

\end{document}

%% file: sm1.tex
\begin{center}
    {\Large \textsc{
    Supplementary material for
    \\
    ``Hydrodynamic fluctuations of stochastic charged cellular automata''
    }}

\vspace{1cm}

{\large Takato Yoshimura$^{1,2}$}

\vspace{0.2cm}
$^1$All Souls College, Oxford OX1 4AL, U.K.\\
$^2$Rudolf Peierls Centre for Theoretical Physics, University of Oxford,
1 Keble Road, Oxford OX1 3NP, U.K.

{\large \v{Z}iga Krajnik$^{3}$}

\vspace{0.2cm}
$^3$Department of Physics, New York University, 726 Broadway, New York, NY 10003, USA

\end{center}

\section{Maximum entropy ensembles and thermodynamics} 
Maximum entropy measures with specified averages of left/right movers and charges factorize
\begin{equation}
	\mathbb{P}({\bf s}) = \prod_{x=1}^{L/2}p_+(s_{2x}) p_-(s_{2x+1}), \label{measure_def}
\end{equation}
in terms of normalized one-site measures
\begin{equation}
	\quad p_-(\pm) = \rho_- \frac{1\pm b}{2}, \quad \quad p_+(\pm) = \rho_+ \frac{1\pm b}{2}, \qquad p_\pm(\emptyset) = \overline \rho_\pm, \label{lrc_measure}
\end{equation}
with $0 \leq \rho_\pm \leq 1$ and $\overline \rho_\pm \equiv 1 - \rho_\pm$ the densities of right/left movers and vacancies respectively. The measure \eqref{measure_def} is an invariant measure for the many-body dynamics, $\mathbb{P}({\bf s}) = \mathbb{P}(\Phi^{\rm full}({\bf s}))$. It can be alternatively parametrized as a grand-canonical Gibbs ensemble
\begin{equation}
	\mathbb{P}({\bf s}) = \mathcal{Z}^{-1} \exp \left[\beta^i \hat Q_i({\bf s})\right], \label{lrc_Gibbs}
\end{equation}
with the index set $\mathcal{I} = \{+, -, c\}$. The conserved charges $\hat Q_i({\bf s}) = \hat Q_i(\Phi^{\rm full}({\bf s}))$ are given as sums of local densities $\hat q^i_{x, t}$
\begin{equation}
	\hat Q^i({\bf s}) = \sum_{x=1}^{L} \hat q^{i}_{x, t}({\bf s}), \label{Q_def}
\end{equation}
with $\hat q^i_{x, t}({\bf s}) = \hat q^i_{x, t}(s_x^t)$ and
\begin{align}
\hat  q^i_{x,t}(\emptyset) &= 0, \qquad \enspace
\hat  q^+_{x,t}(\pm) =
\begin{cases}
1, &{x+t} \in 2\mathbb{Z},\\
0, &{x+t} \in 2\mathbb{Z}+1,\\
\end{cases},\\
\hat q^c_{x,t}(\pm) &= \pm 1, \qquad 
\hat q^-_{x, t}(\pm) = 
\begin{cases}
	0, &{x+t} \in 2\mathbb{Z},\\
	1, &{x+t} \in 2\mathbb{Z}+1,\\
\end{cases}.
\end{align}
Local densities of conserved quantities satisfy a discrete continuity equation of the form
\begin{equation}
	\frac{1}{2}\left(\hat q^i_{x, t+2} - \hat q^i_{x, t}\right) + \hat \jmath^i_{x+1, t+1} - \hat \jmath^i_{x, t+1} = 0, \label{local_continuity}
\end{equation}
with local current densities
\begin{equation}
	\hat \jmath^i_{x, t+1} = \frac{(-1)^x}{2} \left(\hat q^i_{x, t+1+(-1)^x}  -  \hat q^i_{x, t+1} \right).
\end{equation}
The partition function $\mathcal{Z} = \sum_{\bf s} \exp \left[\beta^i \hat Q_i({\bf s}) \right]$ in Eq.~\eqref{lrc_Gibbs} evaluates to 
\begin{equation}
	\mathcal{Z} = \mathcal{Z}_{-}^{L/2} \mathcal{Z}_{+}^{L/2},
\end{equation}
where $\mathcal{Z}_\pm = 1 + 2e^{-\beta_\pm}\cosh \beta_c.$
The corresponding equilibrium free energy per lattice site $f_{\rm eq} = L^{-1}\log \mathcal{Z}$ reads
\begin{equation}
	f_{\rm eq}(\beta_-, \beta_+, \beta_c) = \frac{f_-+ f_+}{2},
\end{equation}
with $f_\pm = 	 \log \left[1 + 2e^{\beta_\pm}\cosh \beta_c \right]$.
The average of an observable $o({\bf s})$ with respect to the measure \eqref{measure_def} is defined as
\begin{equation}
	\langle o \rangle = \sum_{{\bf s}} \mathbb{P}({\bf s})\, o({\bf s}). \label{avg_def}
\end{equation}
Chemical potentials of the exponential measure \eqref{lrc_Gibbs} are related to parameters of \eqref{lrc_measure} by matching averages
\begin{equation}
	L^{-1} \langle Q_\pm \rangle = \partial_{\beta_\pm} f_{\rm eq} = \rho_\pm/2, \qquad  L^{-1}\langle Q_c \rangle = \partial_{\beta_c} f_{\rm eq} = (\rho_- + \rho_+)b/2,
\end{equation}
which gives the identification
\begin{equation}
	\beta_\pm = \log \left[\frac{\rho_\pm \sqrt{1-b^2}}{2\overline \rho_\pm}\right], \qquad \beta_c = \frac{1}{2} \log \left[ \frac{1+b}{1-b} \right],
\end{equation}
with the inverses
\begin{equation}
	\rho_\pm = \frac{\cosh \beta_c}{ \cosh \beta_c  +  e^{-\beta_\pm}/2}, \qquad b =  \tanh \beta_c. \label{beta_to_rhob}
\end{equation}

\section{Dynamical structure factor}
We are interested in the dynamical charge-charge structure factor of the stochastic cellular automaton which we define as
\begin{equation}
S(2x, 2t) = \langle q^c_{2x, 2t} q^c_{0,0}\rangle^c, \label{S_def}
\end{equation}
where $\langle XY \rangle^c =  \langle X Y \rangle - \langle X \rangle \langle Y \rangle $ and  we have  introduced the two-site charge for convenience
\begin{equation}
q^c_{2x, 2t} = \frac{1}{2} [\hat q^c_{2x+1,2t} + \hat q^c_{2x,2t}] \label{c_def}
\end{equation}
to account for the even-odd lattice staggering.

\subsection{Duality}
We compute the structure factor using duality \cite{Schutz_1997}, which maps the problem of computing an $(n+1)$-point function in a stationary measure to computing the dynamics of $n$ particles. Presently, duality is straightforwardly realized by embedding the averaging over strings ${\bf s}$ with respect to the equilibrium probability distribution as
\begin{equation}
\langle o \rangle = \sum_{\bf s}\mathbb{P}({\bf s}) o({\bf s}) = \langle \mathds{1}| o | \varrho \rangle 
\end{equation}
where $|\mathds{1}\rangle = |1 \rangle^{\otimes L}$  is an unnormalized flat state with $|1\rangle = (1, 1, 1)^T$ while $| \varrho \rangle =  \left(|\rho_+\rangle \otimes |\rho_- \rangle\right)^{\otimes L/2}$ is the initial many-body measure which factorizes in terms of one-body measures  with $ |\rho_\pm\rangle = \left(\overline \rho_\pm, \rho_\pm \frac{1+b}{2},  \rho_\pm \frac{1-b}{2}\right)^T$ where we have ordered the local basis as $\{ \emptyset, +, - \}$. The embedding of the local charge reads $\hat q^c = {\rm diag}(0, 1, -1)$ where we abuse notation by using the same symbol for an observable and its embedding. In terms of the above embedding the dynamical structure factor \eqref{S_def} reads
\begin{equation}
    S(2x,2t) = \langle \mathds{1}| q^c_{2x,2t} q^c_{0,0} | \varrho \rangle - \langle \mathds{1}| q^c_{2x,2t} | \varrho \rangle \langle \mathds{1}| q^c_{0,0} | \varrho \rangle. \label{S_embedd}
\end{equation}
In what follows we evaluate the above expression by noting that, once the dynamics of vacancies are eliminated, the evolution of the charge operator in the first term on the right-hand side of Eq.~\eqref{S_embedd} reduces to a one particle problem in the stochastic six-vertex model since it acts on an arbitrary pair of states $|\alpha \rangle = (\alpha, 1-\alpha)^T$ in the $\{+,-\}$ subspace as
\begin{equation}
    U |\alpha \rangle \otimes |\beta\rangle = \overline \Gamma |\alpha \rangle \otimes |\beta\rangle + \Gamma |\beta \rangle \otimes |\alpha\rangle
\end{equation}
For simplicity we hencefort work in the thermodynamic limit $L \to \infty$ and specialize to $\rho_\pm = \rho$.

\subsection{Vacancy elimination}
After eliminating vacancy vertices in the first term on the right-hand side of Eq.~\eqref{S_embedd}, we are left with reduced correlation function which are expressed as sums of one-particle transition probabilities of the symmetric six-vertex model. Introducing $\nu = \overline \rho/\rho$ and carefully counting vacancy combinatorics gives 
\begin{align}
&\langle \mathds{1}| \hat q^c_{2x,2t}\hat q^c_{0,0} |\varrho \rangle \\
&= \rho^{2t+1} \sum_{n_-=0}^{t+x} \sum_{n_+=0}^{t-x-1} \binom{t+x}{n_-} \binom{t-x-1}{n_+} \nu^{n_-+n_+} \left[w\left(t-x-n_+, t+x-n_- \right) (1-b^2) + b^2 \right] \nonumber \\
&= \rho^{2t+1} (1-b^2) \sum_{n_-=0}^{t+x} \sum_{n_+=0}^{t-x-1} \binom{t+x}{n_-} \binom{t-x-1}{n_+} \nu^{n_-+n_+} 
w\left(t-x-n_+,t+x-n_- \right) + \rho^2 b^2 + \rho \overline \rho b^2 \delta_{x,t},  \nonumber\\
&\langle \mathds{1}| \hat q^c_{2x,2t}\hat q^c_{1,0} |\varrho \rangle \\ &=  \rho^{2t+1} \sum_{n_-=0}^{t-x} \sum_{n_+=0}^{t+x-1} \binom{t-x}{n_-} \binom{t+x-1}{n_+} \nu^{n_-+n_+}\left[ w\left( t+x-1/2-n_+, t-x+1/2-n_- \right)(1-b^2) + b^2\right] \nonumber\\
&=\rho^{2t+1}(1-b^2) \sum_{n_-=0}^{t-x} \sum_{n_+=0}^{t+x-1} \binom{t-x}{n_-} \binom{t+x-1}{n_+} \nu^{n_-+n_+}w\left(t+x-1/2-n_+,t-x+1/2-n_-\right)(1-b^2) + \rho^2 b^2,  \nonumber\\
&\langle \mathds{1}| \hat q^c_{2x+1,2t}\hat q^c_{0,0} |\varrho \rangle \\
 &= \rho^{2t+1} \sum_{n_-=0}^{t+x} \sum_{n_+=0}^{t-x-1} \binom{t+x}{n_-} \binom{t-x-1}{n_+} \nu^{n_-+n_+} \left[w\left(t-x-1/2-n_+,t+x+1/2-n_- \right) (1-b^2) + b^2 \right] \nonumber\\
&= \rho^{2t+1} (1-b^2)\sum_{n_-=0}^{t+x} \sum_{n_+=0}^{t-x-1} \binom{t+x}{n_-} \binom{t-x-1}{n_+} \nu^{n_-+n_+} w\left(t-x-1/2-n_+, t+x+1/2-n_- \right) (1-b^2) + \rho^2 b^2,  \nonumber\\
&\langle \mathds{1}| \hat q^c_{2x+1,2t}\hat q^c_{1,0} |\varrho \rangle\\
&= \rho^{2t+1} \sum_{n_-=0}^{t-x} \sum_{n_+=0}^{t+x-1} \binom{t-x}{n_-} \binom{t+x-1}{n_+} \nu^{n_-+n_+}\left[ w\left(t+x-n_+, t-x-n_- \right)(1-b^2) + b^2\right] \nonumber\\
&=\rho^{2t+1}(1-b^2) \sum_{n_-=0}^{t-x} \sum_{n_+=0}^{t+x-1} \binom{t-x}{n_-} \binom{t+x-1}{n_+} \nu^{n_-+n_+}w\left(t+x-n_+, t-x-n_-\right)(1-b^2) + \rho^2 b^2 + \rho \overline \rho b^2 \delta_{x,-t}.  \nonumber
\end{align}
where $w(x, t)$ is the probability of a particle starting from $(0,0)$ to end at position $x$ at time $t$ as we explain in more detail in the next subsection.

\noindent Returning to Eq.~\eqref{S_embedd} and combining the above results we have
\begin{equation}
S(2x, 2t) = \frac{1}{4} \rho \overline \rho  b^2 \left(\delta_{x, t} + \delta_{x, -t} \right) + \frac{1}{4}(1-b^2)\rho [f(2x, 2t) + f(-2x, 2t)]. \label{S_discrete}
\end{equation}
where we have introduced
\begin{align}
f(2x, 2t) = &\rho^{2t}  \sum_{n_-=0}^{t+x} \sum_{n_+=0}^{t-x-1} \binom{t+x}{n_-} \binom{t-x-1}{n_+} \nu^{n_-+n_+} \nonumber \\
&[w\left(t-x-n_+, t+x-n_- \right) + w\left(t-x-1/2-n_+, t+x+1/2-n_- \right)]. \label{f_def}
\end{align}

\subsection{One-particle transition probability}
An expression for the one-particle transition probability from position $y$ to $x$ in $t$ time steps of the symmetric six-vertex model follows by specializing theorem 4.9 of Ref.~\cite{Borodin_Corwin_Gorin_2016} to $N=1$
\begin{equation}
\mathcal{T}^t(y \mapsto x ) = \oint_{c_r} \frac{\dd z_i}{2\pi \ii z}  z^{t+y-x}h^t(z), \label{1particle}
\end{equation}
where $c_r$ is a small positively-oriented contour around the origin and $h$ is a discrete dispersion
\begin{equation}
h(z) = \frac{1+\Gamma(z^{-1}-2)}{1-\Gamma z}.
\end{equation}
We then define the probability distribution $w(x,t)$ for $(x, t) \in \mathbb{Z} \times \mathbb{Z}_{\geq 0}$ as
\begin{align}
w(x, t) = \mathcal{T}^t(0 \mapsto x ) &= \oint_{c_r} \frac{\dd z_i}{2\pi \ii z}  z^{t-x}h^t(z),\\
w(x+1/2, t+1/2) &= \overline \Gamma \sum_{y=0}^x \mathcal{T}^t(0\mapsto y) \Gamma^{x-y} \\
&= \overline \Gamma\oint_{c_r} \frac{\dd z_i}{2\pi \ii z}  z^{t-x}h^t(z) \tfrac{1- (\Gamma z)^{x+1}}{1-\Gamma z} = \overline \Gamma\oint_{c_r} \frac{\dd z_i}{2\pi \ii z}  \frac{z^{t-x}}{1-\Gamma z}h^t(z). \label{w2_int}
\end{align}
where the last equality follows by considering the residues at $z=0$. We now evaluate the integral representation of the one-particle transition probabilities and resum the resulting expression following the method of Ref.~\cite{krajnik2025}. We start by observing that
\begin{equation}
1 + \Gamma(z^{-1}-2) = \frac{(2\Gamma - 1)(1-\Gamma z)+\overline \Gamma^2}{\Gamma z}
\end{equation}
and accordingly expand
\begin{align}
w(x, t) &= \oint_{c_r} \frac{\dd z_i}{2\pi \ii z}  z^{t-x}h^t(z) = \oint_{c_r} \frac{\dd z_i}{2\pi \ii z} \Gamma^{-t}  z^{-x}\frac{\left[(2\Gamma - 1)(1-\Gamma z)+\overline \Gamma^2\right]^t}{(1-\Gamma z)^t}\\
&= \oint_{c_r} \frac{\dd z_i}{2\pi \ii z}  \sum_{p=0}^t \binom{t}{p} \Gamma^{-t} z^{-x} \overline \Gamma^{2p} (2\Gamma -1)^{t-p} (1-\Gamma z)^{-(p+1)}(1-\Gamma z))\\
&=   \sum_{p=0}^t \sum_{u=0}^{\infty} \binom{t}{p} \binom{p+u}{u} \Gamma^{u-t} \overline \Gamma^{2p} (2\Gamma -1)^{t-p} \oint_{c_r} \frac{\dd z_i}{2\pi \ii z} (z^{u-x}-\Gamma z^{u+1-x}) \\
&=   \sum_{p=0}^t \sum_{u=0}^{\infty} \binom{t}{p} \binom{p+u}{u} \Gamma^{u-t} \overline \Gamma^{2p} (2\Gamma -1)^{t-p} (\delta_{u, x}-\Gamma \delta_{u+1, x})\\
&=   \Gamma^{x-t} \sum_{p=0}^t \binom{t}{p} \binom{p+x-1}{x} \overline \Gamma^{2p} (2\Gamma -1)^{t-p}.
\end{align}
Using the Chu-Vandermonde identity $\binom{n+m}{k} = \sum_{j=0}^{k} \binom{n}{j}\binom{m}{k-j}$, we extract and resum the $p$ term in the second binomia in order to get rid of the $2\Gamma -1$ term in the sum
\begin{align}
w(x, t) = &\Gamma^{x-t}  \sum_{p=0}^t \binom{t}{p} \binom{p+x-1}{x} \overline \Gamma^{2p} (2\Gamma -1)^{t-p} =\\
&\Gamma^{x-t} \sum_{p=0}^t \sum_{j=0}^{x} \binom{t}{p} \binom{p}{j} \binom{x-1}{x-j} \overline \Gamma^{2p} (2\Gamma -1)^{t-p} =\\
&\Gamma^{x-t}  \sum_{p=0}^t \sum_{j=0}^{x} \binom{t}{j} \binom{t-j}{p-j} \binom{x-1}{x-j} \overline \Gamma^{2p} (2\Gamma -1)^{t-p} = \\
& \Gamma^{x-t} \sum_{j=0}^{x} \binom{t}{j}  \binom{x-1}{j-1}  \overline \Gamma^{2j} \sum_{p=j}^t \binom{t-j}{p-j}  \overline \Gamma^{2(p-j)} (2\Gamma -1)^{t-p} = \\
& \Gamma^{x+t} \sum_{j=0}^{x} \binom{t}{j}  \binom{x-1}{j-1}  (\overline \Gamma/\Gamma)^{2j}. \label{wes}
\end{align}
Following an analogous procedure we also have 
\begin{align}
w(x+1/2, t+1/2)
&=  \overline \Gamma \sum_{p=0}^t \binom{t}{p} \binom{p+x}{x} \Gamma^{x-t} \overline \Gamma^{2p} (2\Gamma -1)^{t-p}.
\end{align}
Resumming the $p$ term in the second binomial we find
\begin{equation}
w(x+1/2, t+1/2) = \overline \Gamma \Gamma^{x+t} \sum_{j=0}^{x} \binom{t}{j}  \binom{x}{j}  (\overline \Gamma/\Gamma)^{2j}. \label{wos}
\end{equation}

\subsection{Structure factor asymptotics}
To obtain the asymptotic form of the structure factor \eqref{S_discrete} it remains to analyze the large space-time behaviour of the function \eqref{f_def}. Denoting $\gamma = \Gamma/\overline \Gamma$ we have
\begin{align}
&f(2x, 2t) = (\Gamma \rho)^{2t}  \sum_{n_-=0}^{t+x} \sum_{n_+=0}^{t-x-1} \binom{t+x}{n_-} \binom{t-x-1}{n_+} (\nu/\Gamma)^{n_-+n_+} \times \nonumber \\
&\left(\sum_{j=0}^{t-x-n_+} \binom{t+x-n_-}{j}\binom{t-x-1-n_+}{j-1}\gamma^{-2j}+\gamma^{-1}\sum_{j=0}^{t-x-1-n_+} \binom{t+x-n_-}{j}\binom{t-x-1-n_+}{j}\gamma^{-2j}\right) \nonumber\\
&\simeq \sum_{n_-=0}^{t+x} \sum_{n_+=0}^{t-x}\binom{t+x}{n_-} \overline \rho^{n_-}\rho^{t+x-n_-} \binom{t-x}{n_+} \overline \rho^{n_+}\rho^{t-x-n_+} \sum_{j=0}^{t-x-n_+}\frac{j + (t-x-n_+ -j)/\gamma}{t-x}  \times \nonumber \\
&\quad \binom{t+x-n_-}{j} \overline \Gamma^j \Gamma^{t+x -n_- -j} \binom{t-x-n_+}{j}\overline \Gamma^j \Gamma^{t-x -n_- -j} .
\end{align}
Note that the only approximation we have used is in the second to last line by changing the summation boundaries by one. We are now in a position to approximate the expression using the de Moivre-Laplace theorem
\begin{equation}
\binom{t}{n}p^n q^{t-n} \simeq \frac{1}{\sqrt{2\pi t p q}} e^{-\frac{(n-pt)^2}{2t pq}}, \qquad p+q=1,
\end{equation}
which gives
\begin{align}
&f(2x, 2t) \\
&\simeq \sum_{n_-=0}^{t+x} \sum_{n_+=0}^{t-x} \sum_{j=0}^{t-x-n_+}  \frac{j + (t-x-n_+ -j)/\gamma}{t-x}  
\frac{1}
{(2\pi)^2 \rho \overline \rho \Gamma \overline \Gamma \sqrt{t^2-x^2} \sqrt{(t+x-n_-)(t-x+n_+)}} \nonumber\\
&\exp\left[
-\frac{(n_- - (t+x)\overline \rho)^2}{2(t+x)\rho \overline \rho} 
-\frac{(n_+ - (t-x)\overline \rho)^2}{2(t-x)\rho \overline \rho}
-\frac{(j - (t+x-n_-)\overline \Gamma)^2}{2(t+x-n_-)\Gamma \overline \Gamma}
-\frac{(j - (t-x-n_+)\overline \Gamma)^2}{2(t-x-n_+)\Gamma \overline \Gamma}
 \right]. \nonumber
\end{align}
It is straightforward to see that to leading order for $t\gg x$ the exponential has a unique maximum
\begin{equation}
n^*_\pm = t \overline \rho, \qquad j^* = t \rho \overline \Gamma,
\end{equation}
where it vanishes.
Anticipating diffusive asymtptotics we accordingly introduce the variables
\begin{equation}
x = \xi t^{1/2}, \qquad n_\pm = \overline \rho t + s_\pm t^{1/2}, \qquad j = \rho \overline \Gamma t + y t^{1/2},
\end{equation}
whereupon we find for large $t$
\begin{equation}
f(2x, 2t) \simeq \frac{1}{(2\pi)^2 \rho \overline \rho \Gamma t^{1/2}} \iint_{\mathbb{R}^2} \dd s_- \dd s_+ \,e^{- G(s_-, s_+)}  \int_{\mathbb{R}}  \dd y\,e^{-R(s_-, s_+, y)}
 \label{integral}
 \end{equation}
with 
\begin{align}
G(s_-, s_+) &= \frac{(s_+ + \overline \rho \xi )^2+ (s_- - \overline \rho \xi )^2}{2\rho \overline \rho},\\
R(s_-, s_+, y) &= \frac{(y + \overline \Gamma(s_- - \xi))^2+ (y + \overline \Gamma(s_- + \xi))^2}{2\rho \Gamma \overline \Gamma}.
\end{align}
The integral over $y$ gives
\begin{equation}
\int_{\mathbb{R}}  \dd y\,e^{-R(s_-, s_+, y)} = \sqrt{\pi \Gamma \overline \Gamma \rho}\ e^{-\frac{1}{4\gamma \rho}(s_+ - s_- + 2\xi)^2}.
\end{equation}
Also carrying out the integrals over $s_\pm$ we find
\begin{equation}
\iint_{\mathbb{R}^2} \dd s_- \dd s_+ \,e^{- G(s_-, s_+)}  \int_{\mathbb{R}}  \dd y\,e^{-R(s_-, s_+, y)} = 2 \pi^{2} \Gamma \rho \overline \rho   \frac{e^{- \xi^2/D}}{\sqrt{\pi D}} ,
\end{equation}
where 
\begin{equation}
D  = \frac{1- \overline  \Gamma \rho}{\overline \Gamma \rho}. \label{sigma_def}
\end{equation}
The diffusive scale $ x \sim t^{1/2}$ asymptotics of the function $f$ are then
\begin{equation}
f(x, t) \simeq  \frac{e^{-x^2/2D t}}{\sqrt{2\pi D t}}.
\end{equation}
Finally, returning back to Eq.~\eqref{S_discrete} we have
\begin{equation}
S(x, t) \simeq \frac{1}{2} \rho \overline \rho  b^2 \left[\delta(x-t)+ \delta(x+t) \right] +\rho(1-b^2) \frac{e^{-x^2/2D t}}{\sqrt{2\pi D t}}, \label{S_hydro}
\end{equation}
with $D$  defined in Eq.~\eqref{sigma_def}. The notation in Eq.~\eqref{S_hydro} is somewhat imprecise - the first term holds on the ballistic scale $x\sim t$ while the second describes the diffusive scale $x \sim t^{1/2}$. We note that the result \eqref{S_hydro} is a vacancy-renormalized generalization of a similar result in Section III of the Supplementary Material of Ref.~\cite{Kos2021} which is recovered for $\rho=1$.


%% file: main.bbl
\begin{thebibliography}{47}
\bibitem [{sel()}]{self_coupling}%
  \BibitemOpen
  \href@noop {} {}\bibinfo {note} {This can be readily seen by noting that $\p v_i^\eff/\p n_i=G_i^{~ii}=0$, where $G_i^{~jk}$ is the so-called G-matrix that controls the strength of coupling among normal modes. See also Ref.~\cite{Hubner2025}.}\BibitemShut {Stop}%
\bibitem [{Doy()}]{Doyon_noise_2025}%
  \BibitemOpen
  \href@noop {} {}\bibinfo {note} {B. Doyon, Hydrodynamic noise in one dimension: projected Kubo formula and its vanishing in integrable models}\BibitemShut {NoStop}%
\end{thebibliography}
